\documentclass[letter,twocolumn]{jpsj3}
\usepackage{txfonts,braket,color,ulem,bm}
\usepackage{booktabs, multirow} %表のパッケージ

\title{Nonrelativistic Piezomagnetic Effect in an Organic Altermagnet}

\author{Makoto Naka$^1$\thanks{m-naka@mail.dendai.ac.jp}, 
Yukitoshi Motome$^2$\thanks{motome@ap.t.u-tokyo.ac.jp}, 
Tsuyoshi Miyazaki$^3$\thanks{MIYAZAKI.Tsuyoshi@nims.go.jp},
and 
Hitoshi Seo$^4$\thanks{seo@riken.jp}}
\inst{$^1$School of Science and Engineering, Tokyo Denki University, Ishizaka, Saitama 350-0394, Japan \\
$^2$Department of Applied Physics, The University of Tokyo, Bunkyo, Tokyo 113-8656, Japan \\
$^3$Research Center for Materials Nanoarchitectonics~(WPI-MANA),
National Institute for Materials Science, Tsukuba, Ibaraki 305-0044, Japan \\
$^4$RIKEN Center for Emergent Matter Science, Wako, Saitama 351-0198, Japan}

\abst{We theoretically study the piezomagnetic effect on the altermagnetic state in 
$\kappa$-type molecular conductors, focusing on its nonrelativistic mechanism. 
By introducing shear stress as a monoclinic distortion, we evaluate variations
 in the effective tight-binding model using first-principles calculations. 
Using the derived parameters, we investigate the Hubbard model and its effective Heisenberg model 
on the two-dimensional (distorted) $\kappa$-type lattice within mean-field approximation. 
We show that the system exhibits the piezomagnetic effect, i.e., a net magnetization induced at finite temperatures 
 in the undoped insulating state and both in the ground state and at finite temperatures upon doping.
 In a real-space picture, this uniform magnetization arises from the ferrimagnetic spin structure due to inequivalent spin sites induced by lattice distortion. 
 Meanwhile, in a momentum-space picture, it stems from the {\it s}-wave spin splitting of the electron and magnon bands, independent of spin-orbit coupling.
We find that this nonrelativistic piezomagnetism remains finite, but becomes smaller in the limit of strong dimerization 
 where the energy gap between the bonding and antibonding orbitals is infinitely large and the {\it d}-wave altermagnetic spin splitting is absent, 
 highlighting the importance of the multi-orbital nature.}

\begin{document}
\maketitle

%\section{Introduction}

Organic conductors, which have been studied 
 as typical strongly correlated electron systems~\cite{Lebed,Seo_ChemRev,Ardavan}, 
 have recently been revealed to exhibit peculiar properties in their antiferromagnetic (AFM) states. 
Specifically, collinear AFM ordering on the double herringbone arrangement of molecules, called $\kappa$-type,
 was proposed to show nonrelativistic spin splitting (SS) in the electronic and magnon bands 
 and a spin current generation under electric field or thermal gradient~\cite{Naka_spincurrent}.  
Model calculations explicitly demonstrated that the anisotropy of electron hoppings 
 dependent on sublattices is the origin of these phenomena.
Furthermore, this AFM state breaks macroscopic time reversal symmetry 
 and then the anomalous Hall effect can be expected when the spin-orbit coupling is considered~\cite{Naka_AHE}.  
These features---the nonrelativistic SS,~\cite{Noda,Okugawa,Hayami,Yuan,Hayami2} 
 its resultant spin current conductivity~\cite{Naka_spincurrent,Ahn,Naka_perovskite1,Gonzalez}, 
 and the anomalous Hall effect 
 sometimes discussed in the finite frequency range~\cite{Naka_AHE,Solovyev,Smejkal_SciAdv,Naka_perovskite2,Sasabe,Hariki}---are in fact the main characteristics of 
 altermagnets~\cite{Smejkal1,Smejkal2,Naka_npj}, 
 which are currently under intensive research. 

Another physical property of altermagnetism, 
 the piezomagnetic effect (PME), has been discussed extensively~\cite{Hayami,Ma,Consoli,McClarty,Yershow,Ogawa,Cheong}, 
 and was recently observed in the inorganic altermagnet MnTe~\cite{Aoyama}. 
The PME is a cross-correlated phenomenon characterized by linear coupling 
 between uniform magnetization and mechanical stress in a magnetic crystal; 
 it has been discussed since early days~\cite{Dzialoshinskii,Moriya} 
 and was observed in AFM ordered CoF$_2$ and MnF$_2$~\cite{Borovik-Romanov}.
 These compounds have rutile-type structures, similar to RuO$_2$, 
 an intesively studied altermagnet candidate, 
 and their AFM order breaks macroscopic time-reversal symmetry. 
 As the microscopic mechanism of this effect, the role of spin-orbit coupling was investigated~\cite{Moriya}. 

On the other hand, the nonrelativistic effect in altermagnets  
 may also lead the PME. %as pointed out in Refs.~\citen{Hayami,Ma,Yershow,Ogawa}. 
For example, in Refs.~\citen{Hayami} and~\citen{McClarty}, the coupling bewteen uniform magnetization and a strain field
 has been pointed out to exist even without the spin-orbit coupling, from the symmetry point of view. 
Such a nonrelativistic PME has been demonstrated 
 in Ref.~\citen{Ma} for a monolayer of V$_2$Se$_2$O by first-principles calculations.
In addition, in Refs.~\citen{Consoli} and \citen{Yershow},
two-dimensional (2D) Heisenberg models with altermagnetic ground states 
 were shown to exhibit uniform magnetization under strain at finite temperatures ($T$), resulting in a ferrimagnetic spin structure. 
In organic compounds with flexible lattices and weak spin–orbit coupling, nonrelativistic contributions to the 
 effect are expected to be particularly pronounced; 
 however, its nature targeting actual materials has not yet been investigated.

In this Letter, we study the PME in the organic altermagnet candidate $\kappa$-(BEDT-TTF)$_2$Cu[N(CN)$_2$]Cl by combining first-principles evaluation of strain effects and model calculations. 
We first derive the effective tight-binding model using first-principles calculations, and then  analyze the two-dimensional Hubbard model and the effective Heisenberg model within mean-field approximation. 
We show the emergence of a nonrelativistic PME at finite $T$ in the undoped insulating state and both in the ground state and at finite $T$ upon doping, in the presense of the strain.
Besides, we find that SS in the unstrained case is not essential for the PME and underscore the importance of multi-orbital nature in the BEDT-TTF dimer~\cite{Naka_Ishihara} for a sizable response. 
Our results highlight that molecular crystals sensitive to pressure can serve as an ideal platform for exploring the PME.

The first-principles band calculations were performed 
 based on a plane-wave density-functional theory 
 within the generalized gradient approximation~\cite{Perdew}, utilizing the QUANTUM ESPRESSO code version 7.2~\cite{Giannozzi} with 
 scalar-relativistic pseudopotentials generated by the projected augmented wave 
 formalism~\cite{Blochl}. Maximally localized Wannier functions were
 generated using the WANNIER90 package~\cite{Souza} by setting
 a Wannier center at each BEDT-TTF molecule. 
The cutoff energies for plane waves and charge densities are set to 40 and 200 Ry, respectively, 
 and a Gaussian smearing method was used with 2 $\times$ 1 $\times$ 2 and 4 $\times$ 2 $\times$ 4 uniform {$\bm k$}-point meshes defined in the space group {\it Pnma} during the structural optimization and self-consistent-field calculation, respectively.

%**********************************************************************
 \begin{figure}
%  \vspace*{2em}
  \begin{center}
  \includegraphics[width=1.0\columnwidth, clip]{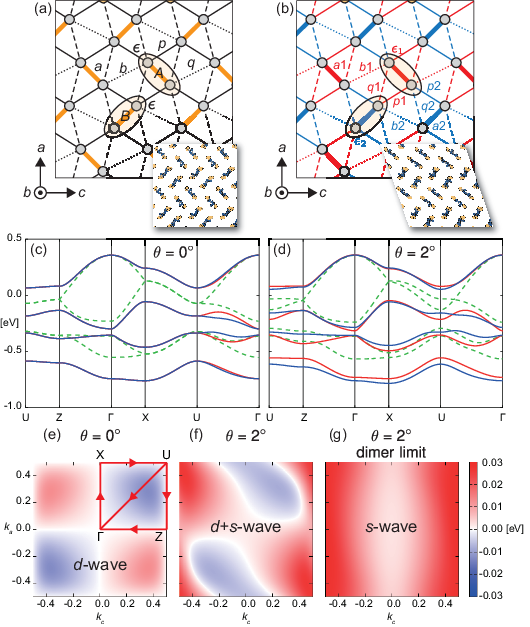}
  \end{center}
  \vspace*{-0.5em}
  \caption{(Color online) Schematic illustrations of the $\kappa$-type lattice structure 
  (a) without and (b) with the monoclinic distortion induced by shear stress. 
  The circles and the ellipses represent the molecular sites and the dimers, respectively. 
  The networks of the bonds are shown: 
  four kinds $a$, $b$, $p$, and $q$ in the undistorted case in (a), 
  and eight kinds \{$a1$, $a2$\}, \{$b1$, $b2$\}, \{$p1$, $p2$\}, and \{$q1$, $q2$\} in the distorted case in (b). 
 Their tight-binding band structures (c) without (tilting angle $\theta=0$) and (d) with the monoclinic distortion ($\theta = 2^\circ$).
 The dashed and solid lines represent those for the paramagnetic ($U=0$) and AFM ($U=1$~eV) phases, 
 where the up and down spin bands are denoted by red and blue, respectively.
 The Fermi energies are set to zero.
 The ${\bm k}$-point path in the orthorhombic Brillouin zone is shown in (e).
 We note that the same symmetry indices are also used in the monoclinic case in (d) for convenience.
 The ${\bm k}$-space distributions of the SS of the top bands in the AFM ground states for (e) $\theta = 0^\circ$, (f) $\theta = 2^\circ$, 
 and (g) the dimer limit for $\theta = 2^\circ$.
 }
  \label{fig1}
  \vspace*{-2.5em}
\end{figure}
%**********************************************************************

As for the effect of shear stress, for simplicity, 
 starting from the experimental crystal structure at ambient pressure for 15~K~\cite{structure}, 
 we tilt the \textit{a}-axis within the \textit{ac}-plane, producing  
 an orthorhombic-to-monoclinic distortion, and while fixing the volume of the unit cell, 
 structurally optimize the internal coordinates. 
This results in the modulation of inter-molecular transfer integrals 
 as shown in Figs.~\ref{fig1}(a) and \ref{fig1}(b), where 
 the 2D network at the ambient pressure orthorhombic and 
 that under the monoclinic distortion, respectivively, are illustrated. 
 
In Table I, 
 we list the transfer integrals by varying the tilting angle $\theta$.
The intradimer $a$, and the interdimer $b, p$, and $q$ bonds in the orthorhombic case [Fig.~\ref{fig1}(a)]
 are modulated as \{$a1$, $a2$\}, \{$b1$, $b2$\}, \{$p1$, $p2$\}, and \{$q1$, $q2$\} [Fig.~\ref{fig1}(b)], respectively.
They vary about 8 -- 20 \% at $\theta=2^\circ$, 
 a typical degree of distortion 
 in organic crystals under pressure of the order of GPa~\cite{Rose,Itoi}. 
The on-site potential energy, $\epsilon_i$, for the two BEDT-TTF molecules that becomes 
 crystallographically inequivalent for $\theta \neq 0$ is also evaluated. 
In Figs.~\ref{fig1}(c) and \ref{fig1}(d), 
 the band structures without strain, $\theta=0^\circ$, and with distortion of $\theta=2^\circ$, 
 are shown. 
Owing to the monoclinic distortion, the band degeneracy along the Brillouin zone boundaries are lifted. 

Now we investigate the PME by 
 calculating the magnetic structures under the variation of these parameters. 
We consider the 2D Hubbard model at three-quarter filling, 
 which has been studied intensively for the undistorted case~\cite{Naka_spincurrent,Kino,Seo_ChemRev}.  
The Hamiltonian is described as 
\begin{align}
{\cal H}_\textrm{Hubb} =\sum_{\langle i,j \rangle, s} t_{ij} \left( c^\dagger_{is} c_{js}^{} + \mathrm{h.c.} \right)
 + \sum_{i} \left( \epsilon_i n_{i} + U n_{i\uparrow} n_{i\downarrow}  \right), 
\end{align}
where $c_{is}$ ($c^\dagger_{is}$) and $n_{is}$ ($= c^\dagger_{is}c_{is}$) are the annihilation (creation)
and number operators of an electron at $i$th molecular site 
with spin $s$, respectively, and $n_i =  n_{i\uparrow}+ n_{i\downarrow}$. 
The transfer integrals $t_{ij}$ are considered for the pairs $\langle i,j \rangle$ 
 along the bonds shown in Figs.~\ref{fig1}(a) and \ref{fig1}(b), as mentioned above. 
$U$ is the on-site Coulomb repulsion treated within the Hartree-Fock approximation. 

\begin{table}
    \centering
    \caption{
    Tight-binding parameters estimated by the first-principles calculation by varying the tilting of the $a$-axis with angle
    $\theta = 0^\circ$, $1^\circ$, and $2^\circ$, in unit of eV.
    ``bond1" and ``bond2" denote the two inequivalent bonds shown in Fig.~\ref{fig1}(b).
    }
%    \vspace{2mm}
    \begin{tabular}{cccccc}
    \toprule
        & \multirow{2}{*}{orthorhombic ($\theta=0^\circ$)} 
        & \multicolumn{2}{c}{monoclinic ($\theta=1^\circ$)} 
        & \multicolumn{2}{c}{monoclinic ($\theta=2^\circ$)}
        \\ \cmidrule{3-6}
        & & bond1 & bond2 & bond1 & bond2 \\
        \midrule
        \midrule
        $\epsilon_i$ & 4.160 & 4.165 & 4.158 & 4.172 & 4.161 \\
        $t_a$ & -0.193 & -0.185 & -0.200 & -0.178 & -0.208 \\
        $t_b$ & -0.068 & -0.071 & -0.065 & -0.075 & -0.068 \\ 
        $t_p$ & -0.097 & -0.106 & -0.090 & -0.116 & -0.082 \\
        $t_q$ &  0.050 & 0.053 & 0.056 & 0.056 & 0.043 \\
    \bottomrule
    \end{tabular}
    \label{table1}
\vspace*{-2.5em}
\end{table}
%**********************************************************************

Without the strain ($\theta=0^\circ$), 
 the on-site Coulomb repulsion leads to an AFM insulating state
 where the sites on the two kinds of dimers {\it A} and {\it B} 
 show opposite spin directions~\cite{Kino}, 
 say, in the {\it z} direction, 
 as $\langle s^z_A \rangle=- \langle s^z_B\rangle$.
 Here $\langle s^z_\alpha \rangle$ ($\alpha = A, B$) 
 represents the expectation value of $z$-spin moment on dimer $\alpha$, 
 while the two sites within each dimer have the same value.  
Since the dimers are connected by glide symmetry, 
 these spins are compensated resulting in no net uniform magnetization, 
 i.e., $m_\textrm{u} \equiv \langle s^z_A \rangle +\langle s^z_B\rangle = 0$. 
 In this state, the {\it d}-wave altermagnetic SS appears~\cite{Naka_spincurrent}, 
 as shown in Figs.~\ref{fig1}(c) and \ref{fig1}(e). 

\begin{figure}
% \vspace*{2em}
 \begin{center}
  \includegraphics[width=0.85\columnwidth, clip]{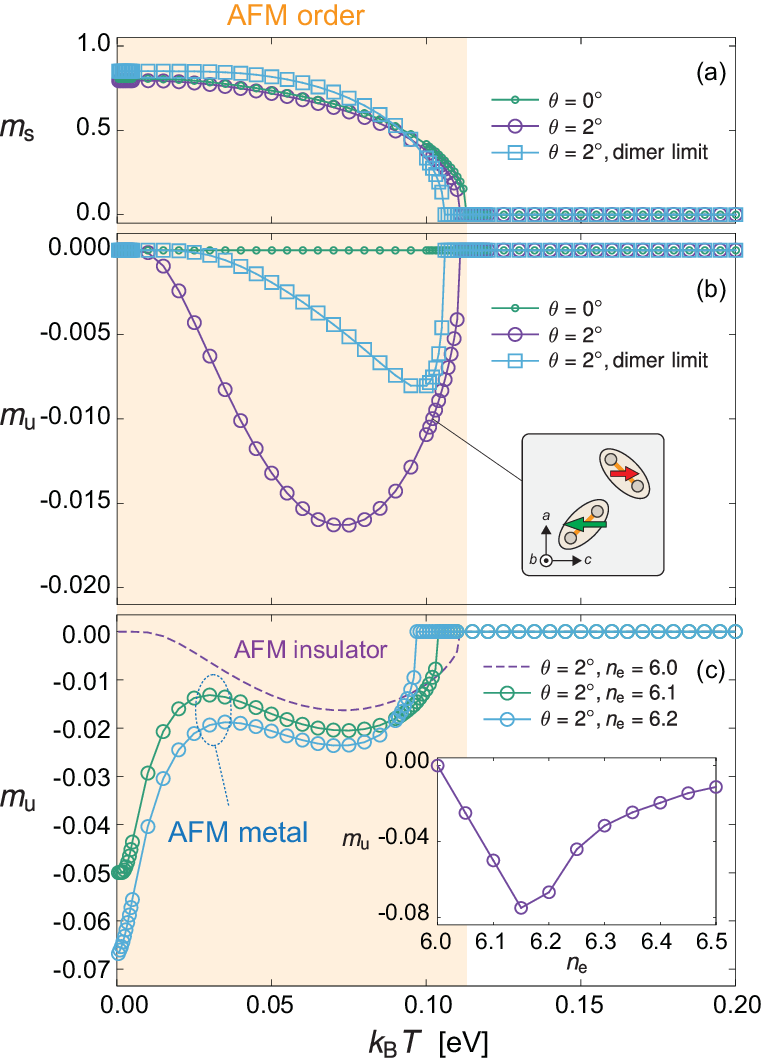}
 \end{center}
 \vspace*{-0.5em}
 \caption{(Color online) $T$ dependences of (a) the staggered magnetization $m_\textrm{s}$ 
 and (b) the net uniform magnetization $m_\textrm{u}$. 
 The undistorted ($\theta=0^\circ$), monoclinically distorted ($\theta=2^\circ$), and its dimer limit (see text) cases are shown. 
 The spin pattern in the ferrimagnetic state is schematically depicted in the inset.
 (c) $T$ dependences of $m_\textrm{u}$ varing the number of electrons in the unit cell $n_{\rm e}$ at $\theta=2^\circ$.
 The inset shows the $n_{\rm e}$ dependence of $m_{\rm u}$ in the ground state at $\theta=2^\circ$.
}
 \label{fig2}
 \vspace*{-2.5em}
\end{figure}

On the other hand, under monoclinic distortion, 
 the dimers {\it A} and {\it B} are no longer equivalent, with no symmetry operation connecting them. 
In Figs.~\ref{fig2}(a) and \ref{fig2}(b), 
 we show the $T$ dependence of staggered and uniform magnetization, 
 $m_\textrm{s} \equiv \langle s^z_A \rangle -\langle s^z_B\rangle$ and $m_\textrm{u}$, 
 respectively, with fixed $U=1$~eV, 
 for $\theta=0^\circ$ and $\theta=2^\circ$.
Under the distortion, in the ground state, 
 the staggered component is almost identical to the undistorted case and there is no net magnetization. 
 %Although the sublattices are inequivalent in this way, 
 Such a state, in which the sublattice magnetizations completely cancel each other 
 even though the sublattices are inequivalent,
 is %often 
 referred to as ``compensated ferrimagnetism'' in recent years~\cite{Mazin,Kawamura,note_compensated}.

However, at finite $T$, $m_\textrm{u}$ is induced by the distortion, 
 stabilizing a ferrimagnetic state as depicted in the inset of Fig.~\ref{fig2}(b). 
 This is the PME.
As $T$ is raised from absolute zero, 
 $m_\textrm{u}$ increases and reaches a maximum in its absolute value, 
 and then decreases down to zero toward $T_{\rm N}$.
 The band structure and SS in the distorted AFM ground state is shown in Figs.~\ref{fig1}(d) and \ref{fig1}(f), respectively; 
 the {\it d}-wave SS is deformed, with an additional extended {\it s}-wave SS being superimposed.
 The difference in the excitation energies across the AFM gap between up- and down-spin electrons, 
 induced by this {\it s}-wave splitting, gives rise to the PME at finite $T$. 
 In contrast, in the ground state without such thermal excitations, the numbers of up- and down-spin electrons are equal, 
 resulting in a fully compensated magnetization.

%**********************************************************************
%**********************************************************************

%**********************************************************************
\begin{figure}
% \vspace*{2em}
 \begin{center}
  \includegraphics[width=0.9\columnwidth, clip]{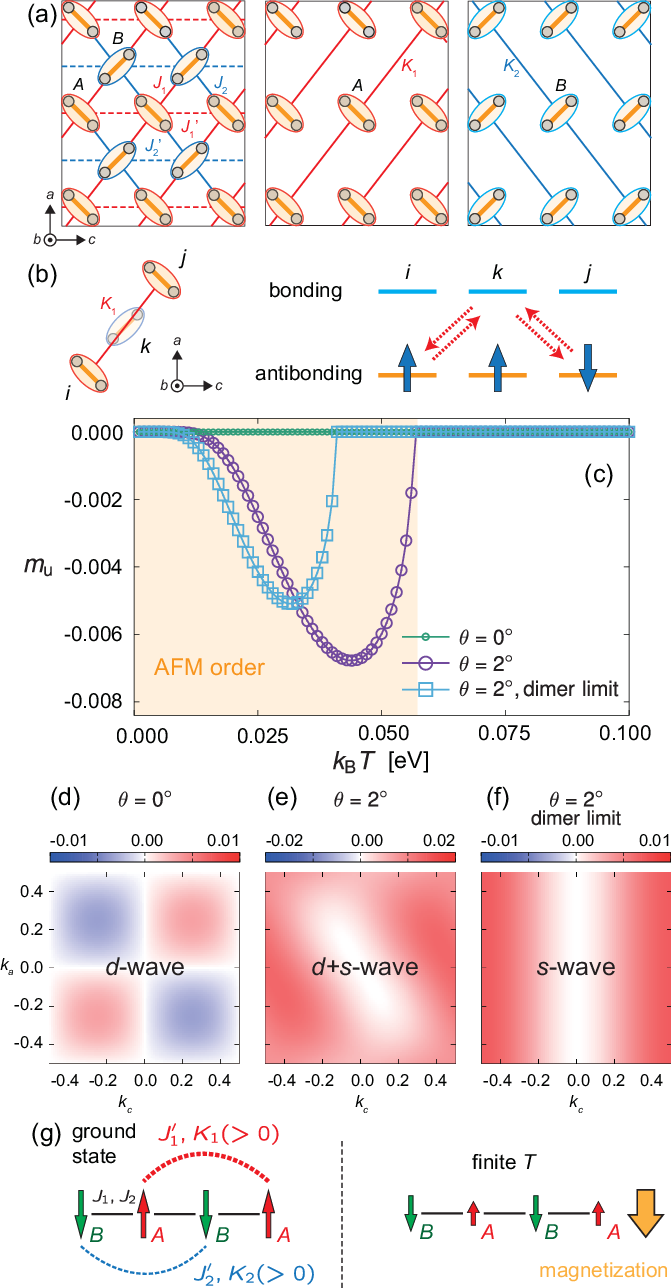}
 \end{center}
 \vspace*{-0.5em}
 \caption{(Color online) 
 The interdimer exchange interactions in the effective Heisenberg model 
 for $\kappa$-type compounds with monoclinic distortion. 
 The NN exhange interactions, $J_1$, $J_2$, $J_1'$, $J_2'$, and the NNN ones, $K_1$, $K_2$, which make the system an altermaget, are shown.
 In the orthorhombic structure without the distortion, $J \equiv J_1=J_2$, $J' \equiv J_1'=J_2'$, and $K \equiv K_1=K_2$ hold.
 (b) Schematic illustration of a perturbation process leading to the NNN exchange interaction in the hole picture, 
  where the transfer integrals between the bonding and antibonding orbitals are essential. 
 (c) $T$ dependences of the net uniform magnetization $m_\textrm{u}$, 
 for the undistorted ($\theta=0^\circ$) and monoclinically distorted ($\theta=2^\circ$) cases, 
 obtained by the mean-field approximation for the Heisenberg model.
 The ${\bm k}$-space distributions of the magnon SS in the AFM ground states for (d) $\theta = 0^\circ$, (e) $\theta = 2^\circ$, 
 and (f) the dimer limit of (e).
 (g) A schematic illustration for the mechanism of the nonrelativistic PME (see text). 
 }
 \label{fig3}
 \vspace*{-2.5em}
\end{figure}
%**********************************************************************

Let us discuss the relationship between SS and the PME, with a focus on the role of dimerization.
We artificially tune the intradimer transfer ingetrals as $t_{a1} \rightarrow t_{a1}-\Delta$ and $t_{a2} \rightarrow t_{a2}-\Delta$; 
the increase in $\Delta$ leads to the separation of bonding and antibonding orbitals of the dimers and for $\Delta = \infty$ the model is reduced to a single-orbital Hubbard model on the anisotropic triangular lattice~\cite{Kino}. 
In this dimer limit, the ${\it d}$-wave SS and the bonding/antibonding orbital degree of freedom are lost both in the orthorhombic and monoclinic cases~\cite{Naka_spincurrent}. 
However, we find that while the uniform component $m_{\rm u}$ becomes smaller as $\Delta$ increase, it converges to a finite value in the dimer limit
 of $\Delta \to \infty$, as plotted in Figs.~\ref{fig2}(a) and \ref{fig2}(b). 
Interestingly, we find that in the dimer limit with monoclinic distortion, 
only the {\it s}-wave component of the spin-splitting survives, as shown in Fig.~\ref{fig1}(g). 
The smaller but finite values of $m_\textrm{u}$ in the dimer limit indicate that 
 even when the {\it d}-wave SS---though symmetry-imposed---is negligibly small, the nonrelativistic PME remains finite and the multi-orbital nature leads to the enhancement of the induced magnetization. 
 
As shown above, the PME vanishes at $T=0$ in the undoped insulating state. 
However, when carriers are doped, 
%and spin-split Fermi surfaces are present
a finite magnetization emerge even at $T=0$~\cite{Ogawa}.
Figure~\ref{fig2}(c) shows the electron doping dependences of the PME at $\theta=2^\circ$, 
 where the number of electrons in the unit cell, $n_\textrm{e}$, is increased from three-quarter filling ($n_{\rm e} = 6.0$). 
 Under the doping, the AFM metallic state is stabilized at low $T$, and the uniform magnetization $m_{\rm u}$ 
 shows the hump below $T_{\rm N}$ as in the undoped case, but it starts to increase again below around $k_{\rm B}T/U = 0.03$.
 The magnitude of $m_{\rm u}$ in the ground state increases with an increase of $n_{\rm e}$, 
 while reaches a peak at $n_{\rm e} \simeq 6.15$ and then decreases as shown in the inset of Fig.~\ref{fig2}(c).
% while $T_{\rm N}$ decreases. 
 The low-$T$ increase in magnetization originates from the SS of the Fermi surfaces, 
 which induces a difference between the up-spin and down spin electron densities.
 The upturn in the $T$ dependence of $m_{\rm u}$ is governed by the amplitude of the {\it s}-wave splitting in the top band, 
 shown in Figs.~\ref{fig1}(f) and \ref{fig1}(g), where the Fermi energy is located under the electron doping.
 
We can furthermore elaborate on the mechanism of the nonrelativistic PME 
 by deriving the effective 2D Heisenberg model through the perturbation expansion with respect to $t_{ij}/U$. 
We follow the treatment of Ref.~\citen{Naka_spincurrent}, 
 which demonstrated the {\it d}-wave SS in the magnon band for the undistorted case. 
The minimal Heisenberg model describing the altermagnetic properties of $\kappa$-type compounds is given by~\cite{Naka_spincurrent} 
\begin{align}
 {\cal H}_{\rm Heis} = \sum_{\langle ij \rangle} J_{ij} \ {\bm S}_{i} \cdot {\bm S}_{j} 
 + \sum_{\langle ij \rangle'} J'_{ij} \ {\bm S}_{i} \cdot {\bm S}_{j} 
 + \sum_{\langle\langle ij \rangle\rangle} K_{ij} \  {\bm S}_{i} \cdot {\bm S}_{j}\ ,  
 %+ K'\sum_{\langle\langle ij \rangle\rangle'} {\bm S}_{i} \cdot {\bm S}_{j}, 
 \label{eq:heis}
 \end{align} 
 where $\langle ij \rangle$ and $\langle ij \rangle'$ stand for 
 the nearest-neighbor (NN) bonds of the isosceles triangular lattice, on the legs in two directions and the bases, respectively; 
 $\langle\langle ij \rangle\rangle$ is the next-nearest-neighbor (NNN) bonds along one of the leg direction, 
 which depends on the sublattices $A$ and $B$ [see Fig.~\ref{fig3}(a)]. 
The NNN exchange interaction originates from the interorbital transfer integrals between the neighboring dimers as illustrated in Fig.~\ref{fig3}(b), 
 leading to the {\it d}-wave spin spliting {mentioned above~\cite{Naka_spincurrent}.
 %, while the other direction is omitten since the values are small [see Fig.~\ref{fig3}(a)].  
 In the presence of the monoclinic distortion, 
 each of these three kinds of bonds splits into two inequivalent ones, as shown in Fig.~\ref{fig3}(a). 
 Consequently, the exchange interactions are modulated as 
 $J_{ij}=J \rightarrow \{J_1,J_2\}$, $J'_{ij}=J' \rightarrow \{J'_1,J'_2\}$, and $K_{ij}=K \rightarrow \{K_1,K_2\}$, 
 where $J$, $J'$, and $K$ are the values for the orthorhomic structure, via the splitting of the transfer integrals listed in Table~\ref{table2}.
Since the monoclinic strain diminishes the glide symmetry existed in the orthorhombic case, 
 the couplings also lose the glide symmetry. 
 
%**********************************************************************
\begin{table}
    \centering
    \caption{
    Exchange paremeters at $U=1$ eV obtained by the perturbation expansion using the transfer integrals in Table~\ref{table1}, for the orthorhombic and monoclinic structures in unit of meV.
    }
%    \vspace{2mm}
    \begin{tabular}{cccccc}
    \toprule
        & \multirow{2}{*}{orthorhombic ($\theta=0^\circ$)} 
        & \multicolumn{2}{c}{monoclinic ($\theta=1^\circ$)} 
        & \multicolumn{2}{c}{monoclinic ($\theta=2^\circ$)}
        \\ \cmidrule{3-6}
        & & bond1 & bond2 & bond1 & bond2 \\
        \midrule
        \midrule
        $J$ & 60.1 & 70.2 & 51.4 & 82.0 & 44.2 \\
        $J'$ & 10.9 &11.9 & 9.98 & 13.3 &  9.37\\ 
        $K$ & 1.65  & 2.27 & 1.22 & 3.09 & 0.89 \\
    \bottomrule
    \end{tabular}
    \label{table2}
\vspace*{-2.5em}
\end{table}
%**********************************************************************

In Fig.~\ref{fig3}(c), we show the tempearature dependences of 
 the net magnetization, $m_\textrm{u}$, for $\theta=0^\circ$ and $\theta=2^\circ$, 
 together with the dimer limit at $\theta=2^\circ$, calculated within the mean-field approximation as in Ref.~\citen{Naka_spincurrent}. 
Similar to the case for the Hubbard model discussed above, 
 $m_\textrm{u}$ appears at finite $T$ with similar $T$ dependence in the undoped case and shows reduction in the dimer limit. 
 
The AFM magnon SS obtained by the linear spin-wave theory for Eq.~(\ref{eq:heis}) are shown in Figs.~\ref{fig3}(d)-\ref{fig3}(f); 
 the SS is defined by subtracting the energy of down-spin magnon from that of up-spin magnon. 
 At $\theta = 2^\circ$, the {\it d}-wave magnon splitting at $\theta=0^\circ$ is superimposed with the extended {\it s}-wave component induced by the monoclinic distortion. 
 In the dimer limit at $\theta = 2^\circ$, the {\it d}-wave splitting vanishes, leaving only the extended {\it s}-wave component.
 From the analysis of the magnon dispersion, we find that the extended {\it s}-wave SS induced by the distortion consists of two components: 
 one proprotional to the diferrence between the NNN exchange interactions $K_1-K_2$, and the other to $J_1'-J_2'$. 
 In the dimer limit, $K_1$ and $K_2$, which are associated with the transfer integrals between the bonding and antibonding orbitals in the neighboring dimers, vanish, 
 whereas $J_1'$ and $J_2'$, determined by those between the antibonding orbitals, remain finite. 
 Consequently, the PME decreases toward the dimer limit but retains a finite value, as shown in Fig.~\ref{fig3}(c). 
 Furthermore, the fact that the SS in the dimer limit becomes purely extended {\it s}-wave is also attributed to the vanishing of $K$, 
 which is responsible for the {\it d}-wave SS.

These considerations lead us to an intuitive understanding of the $T$ dependence of the nonrelativistic PME [Fig.~\ref{fig3}(g)].
The couplings between the two sublattices {\it A} and {\it B} are $\{J_1,J_2\}$ 
 while those for {\it A}-{\it A} and {\it B}-{\it B} are $\{J'_1,K_1\}$ and $\{J'_2,K_2\}$, 
 respectively. 
Owing to the different exchange couplings for the latter intra-sublattice bonds, 
 the finite-$T$ fluctuations become different between the two sublattices, 
 as was discussed in Ref.~\citen{Yershow}. 
 In terms of magnons, this can be interpreted as the emergence of net magnetization due to a difference between the number of thermally excited up-spin and down-spin magnons as a result of the {\it s}-wave SS.
We also note that even for $K_1=K_2=0$ the effect remains finite; 
 considering that $K_{ij}$ was responsible for the {\it d}-wave magnon SS 
 in the undistorted case~\cite{Naka_spincurrent}, 
 the altermagnetic SS is not required here, 
 as in the case of the Hubbard model discussed above. 

Finally, we discuss the relevance of our results to experiments. 
As stressed in the introductory part, organic crystals are tunable by external pressure, 
 and uniaxial pressure has been used to control the properties of correlated electrons. 
Our results, showing the induced uniform magnetization of a few percent compared to the staggered component of the AFM order 
 in a moderate range of shear stress, 
 demonstrate organic systems a promising platform for exploring the PME in altermagnets. 
Another issue is the effect of spin-orbit coupling, 
 which has conventionally regarded as the microscopic origin of piezomagnetism. 
In our case the direction of the net magnetization is along the sublattice magnetization. 
In Ref.~\citen{Moriya}, the piezomagnetism owing to the spin-orbit coupling in CoF$_2$ was 
 also predicted to appear also along the sublattice magnetization. 
However, the former exhibits a maximum just below $T_{\rm N}$ at finite $T$, 
while the latter is expected to increase as $T$ approaches zero. 
In organic compounds with relatively weak spin–orbit coupling, the nonrelativistic contribution is likely to dominate over the relativistic one near $T_{\rm N}$.
This suggests that detailed analysis of the $T$ dependence could allow a clear distinction between these two mechanisms.
Last but not least, we comment on the effect of magnetic domains.
 In the $\kappa$-type lattice, there are two types of altermagnetic domains, namely $\{ s^z_A. s^z_B \} = \{ \uparrow, \downarrow \}$ and $\{ \downarrow, \uparrow \}$, 
 and their coexistence can cancel out the present PME. 
 However, these domains are accompanied by weak ferromagnetic moments along the $a$-axis with opposite directions due to the spin–orbit coupling~\cite{Naka_AHE}. 
 Therefore, field cooling is sufficient to align the domains and observe the PME.

\vspace{5mm}

%%%%%%%%%%%%%%%%%%%%%%%%%%%%%%%%%%%%%%%%%%%%%%%%%%%%%%%%

\acknowledgment

We thank T. Aoyama and K. Ohgushi for fruitful discussions. 
This work is supported by JSPS KAKENHI Grant Numbers, 
No. JP19K03723, JP19K21860, JP20H04463, JP23H01129, JP23K25826, JP23K03333, JP25H00838, JP25H01247, the GIMRT Program of the Institute for Materials Research, Tohoku University, No. 202212-RDKGE-0062, and No. 202312-RDKGE-0062.


\begin{thebibliography}{9}
\bibitem{note_compensated}
We note that the term ``(fully) compensated ferrimagnetism" originally refers to a ferrimagnetic system 
 with $T$-dependent behavior of the two sublattices giving rise to a $T$ range 
 where the net magnetization becomes nearly zero, which is useful for application.
\end{thebibliography}
\end{document}